\documentclass[12pt,reqno]{amsart}
\usepackage{amsmath,amsfonts,amsthm,amssymb,mathabx,dsfont}

\usepackage{amsxtra}

\newtheorem{theorem}{Theorem}

\theoremstyle{definition}

\theoremstyle{remark}

\renewcommand{\k}{\mathbf{k}}
\newcommand{\x}{\mathbf{x}}

\newcommand{\y}{\mathbf{y}}
\newcommand\id{\mathds{I}}

\newcommand{\ph}{\varphi}

\newcommand{\C}{\mathbb{C}}

\renewcommand{\epsilon}{\varepsilon}

\newcommand{\F}{\mathcal{F}}

\renewcommand{\phi}{\varphi}
\newcommand{\R}{\mathbb{R}}

\DeclareMathOperator{\im}{Im}

\DeclareMathOperator{\Tr}{Tr}

\begin{document}

\title[Time-Dependent BdG Equations]{Incompatibility of time-dependent Bogoliubov--de-Gennes and  Ginzburg--Landau equations}

\author[R.L. Frank]{Rupert L. Frank}
\address{Mathematics 253-37, Caltech, Pasadena, CA 91125, USA}
\author[C. Hainzl]{Christian Hainzl} 
\address{Mathematisches Institut, Universit\"at T\"ubingen, Auf der Morgenstelle 10, 72076 T\"ubingen, Germany}
\author[B. Schlein]{Benjamin Schlein}
\address{Institute of Mathematics, University of Z\"urich, Winterthurerstrasse 190, 8057 Z\"urich, Switzerland}
\author[R. Seiringer]{Robert Seiringer}
\address{Institute of Science and Technology Austria (IST Austria), Am Campus 1, 3400 Klosterneuburg, Austria}

\begin{abstract}
We study the time-dependent Bogoliubov--de-Gennes equations  for gen\-eric  translation-invariant fermionic many-body systems. For initial states that are close to  thermal equilibrium states at temperatures near the critical temperature, we show that the magnitude of the order parameter stays approximately constant in time and, in particular, does not follow a time-dependent Ginzburg--Landau equation, which is often employed as a phenomenological description and predicts a decay of the order parameter in time. The full non-linear structure of the equations is necessary to understand this behavior.
\end{abstract}

\maketitle

\section{Introduction}

The Ginzburg--Landau (GL) model \cite{GL} is a paradigm for the phenomenological description of phase transitions in physical systems. In the static case, its relation to the microscopic BCS theory \cite{bcd} was clarified by Gor'kov \cite{Gorkov} (see also  \cite{E} and \cite{degennes} for alternative approaches), and a mathematically rigorous derivation of the GL model from BCS theory was recently given in \cite{FHSS, FHSS2}. 
The present work is concerned with the analogous question in the time-dependent case. Arguments for the validity of a time-dependent GL equation can be found in the literature, starting with 
\cite{StSu,AbrTsu,Schmid} 
in the case of superconductors. These attempts were critically analyzed 
in \cite{GE} where it was argued that the equation can only hold in a gapless regime, however; we refer to \cite{Cyrot} for a thorough discussion. By applying similar arguments in the study of superfluid cold gases, it was stated in \cite{melo,randeria} that the non-linear time-dependent GL equation
applies to such systems for temperatures $T$ slightly above the critical temperature. 

The main message of our present work is that a time-dependent GL equation cannot be derived from the time-dependent Bogoliubov--de-Gennes (BdG) equations near the critical temperature,  in contrast to the static case. 
We shall consider the BdG equations for a translation-invariant system, which can be derived in a standard way applying the BCS approximation, either to the Heisenberg equations of motion for the fermion field operators, or the time-dependent Green's function \cite{Cyrot,gorkov58}. We consider a system which is initially close to a thermal equilibrium state near the critical temperature,  with non-vanishing order parameter. We then show that, in contrast to what would be expected from GL  theory, the order parameter does not decay in time. Interestingly, this is a purely non-linear effect. If, instead, we consider the corresponding linear equation, then the solution indeed decays  exponentially in time, on a time scale that can be calculated via the imaginary part of the corresponding resonance pole, which turns out to be proportional to the inverse of the distance to the critical temperature. The non-linear terms formally contribute a term cubic  in the order parameter, which indeed resembles a GL type equation; however, small denominators close to the Fermi sphere invalidate such formal arguments, and prevent the decay on all time scales.

Our claims are mathematically rigorous and are not based on any ad-hoc approximations; they are confirmed numerically in \cite{HaiSey} in the case of a one-dimensional system with contact interactions. 

The present work can be viewed as a continuation of a recent series of studies of the mathematical properties of the BCS theory of superconductivity and superfluidity \cite{HHSS,FHNS,HS-T_C,FL}.
It is motivated by the current interest concerning the applicability of the theory to cold gases, in particular concerning the BCS--BEC crossover \cite{zwerger-book}. In the BCS regime \cite{Leggett, NRS},  a rigorous proof of the emergence of a static GL  equation close to the critical temperature was given in \cite{FHSS, FHSS2}. 
In the BEC regime, the prediction \cite{randeria, melo, Pieri-Strinati,zwerger-1992} of the emergence of the Gross--Pitaevskii equation in the low-density limit was rigorously established, 
both in the static \cite{HS} and the dynamical case \cite{HSch}. 

\section{Model and Main Result}

The starting point of our analysis is the BCS model \cite{bcd,Leggett,HHSS}. The state of a (translation-invariant, three-dimensional\footnote{Our analysis extends to one- and two-dimensional systems in a straightforward way.} system is described in terms of two quantities, the momentum distribution $
\gamma(\k)=\langle  a^\dagger_\k a_\k \rangle$ and the pair density $\hat\alpha(\k)=\langle a_{\k} a_{-\k}\rangle$, determining the Cooper pair wave-function via Fourier transform as $\alpha(\x-\y) = (2\pi)^{-3/2} \int \hat\alpha(\k) e^{i \k\cdot (\x-\y)} d^3k$.  They can be conveniently combined to the $2\times 2$ matrix 
\[ 
\Gamma (\k)  = \left ( \begin{matrix}  \gamma (\k)  &  \hat{\alpha} (\k) \\ \overline{\hat{\alpha} (\k)} & 1 - \gamma (-\k)  \end{matrix} \right ) \, , 
\]
which satisfies  $0 \leq \Gamma(\k) \leq \id_{\C^2}$ for every $\k$. We suppress spin in our notation; the pair density $\hat\alpha$ is assumed, for simplicity, to be a spin singlet. The equilibrium state  at temperature $T\geq 0$ and chemical potential $\mu$ is determined by minimizing the pressure functional
\begin{equation}\label{freeenergy}
\F(\Gamma)= \int_{\R^3}  \!(k^2-\mu) \gamma (\k) d^3k+\int_{\R^3} |\alpha(\x)|^2 V(\x) d^3x -T
S(\Gamma),
\end{equation}
with entropy  $$S(\Gamma) = - \int_{\R^3} {\rm
Tr}_{\C^2}\left[ \Gamma(\k) \ln \Gamma(\k)\right]d^3k.$$ 
We find it convenient to choose units such that $\hbar = 2m = 1$, with $m$ denoting the particle mass. 
In (\ref{freeenergy}), $V(\x)$ denotes a local two-body interaction potential, as appropriate for the description of superfluid gases. Our results can  easily be generalized to include non-local interactions as well, which effectively arise via interactions through phonons in solids, for instance.

It was shown in \cite{HHSS} that the critical temperature $T_c$ for the model (\ref{freeenergy}) is the unique $T$ for which the operator $K_T (-i\nabla) + V(\x)$ has $0$ as its lowest eigenvalue, where 
\begin{equation*}\label{eq:KTp} 
K_T(\k) = \frac {k^2 - \mu}{\tanh \left(\tfrac {k^2- \mu}{2T} \right)}. 
\end{equation*} 
That is, for $T \geq T_c$ the pressure functional (\ref{freeenergy}) is minimized by the normal state $\Gamma_{\rm n}(\k)$ with $\gamma_{\rm n}(\k) = (1+ e^{(k^2-\mu)/T})^{-1}$ and $\hat\alpha_{\rm n}(\k)\equiv 0$, while for $T< T_c$ the pairing density $\hat\alpha(\k)$ does not vanish identically, showing the transition to a superfluid (or superconducting) phase.  Strictly speaking, this characterization  only applies if $T_c$ is strictly positive, which we shall assume henceforth. We shall also assume that $K_{T_c} (-i\nabla) + V(\x)$ has a unique normalized eigenvector $\alpha_*$ corresponding to the eigenvalue $0$; 
for radial $V(\x)$ this corresponds to the assumption that $\alpha_*$ is an $s$-wave.\footnote{For an analysis of the BCS model without  this non-degeneracy assumption, see \cite{FL}.} Smoothness and decay properties of $\alpha_*$ are analyzed in detail in \cite[App.~A]{FHSS}.

For temperatures $T$ slightly below $T_c$, $\hat\alpha(\k)$ is proportional to $(T_c-T)^{1/2} \alpha_*$ to leading order in $T_c - T$. That is, we can write
\begin{equation}\label{ap}
 \hat{\alpha} (\k) =  \psi \hat{\alpha}_* (\k) + \xi (\k) 
 \end{equation}
with $\psi\in\C$ of order $\sqrt{T_c-T}$, and $\xi(\k) \ll \sqrt{T_c-T}$ for small $T_c-T$. The order parameter $\psi$ is, in fact, determined by minimizing the GL type expression
\begin{equation}\label{eq:GL} 
\mathcal{E}_\text{GL} (\psi) =  C_{\rm GL}(T-T_c) |\psi|^2 +  |\psi|^4
\end{equation}
for a suitable constant $C_{\rm GL}>0$. This follows from the analysis  in \cite{FHSS}, which is actually  more general and not restricted to the translation invariant case considered here.

The time-dependent BCS equation, also known as the BdG equation,  has the form 
\begin{equation}\label{eq:BdG-p} i\partial_t \Gamma_t (\k) = \left[ H_t (\k) , \Gamma_t (\k) \right] 
\end{equation}
with effective Hamiltonian   
\[ H_t (\k) = \left( \begin{matrix} k^2 - \mu &  \Delta_t(\k)    \\ \overline{\Delta_t(\k)}  &  \mu - k^2  \end{matrix} \right) ,
\]
where  $\Delta_t(\k)\! = \! 2 (2\pi)^{-3/2} \int \hat V(\k\!-\!\k') \hat \alpha_t(\k') d^3k'$. It can be derived from the Heisenberg equations of motion for the fermion field operators, applying the same BCS-type approximations as in the static case \cite{andreev,Cyrot,kuemmel}. Alternatively, it also follows from the time-dependent Green's function method introduced in \cite{gorkov58} (see also \cite{parks}). While only certain interaction terms are retained in the BCS approximation, the equation (\ref{eq:BdG-p}) allows for pair-creation and annihilation and hence the superfluid density is in general not a conserved quantity. 
Note that $H_t$ depends itself on $\Gamma_t$ through $\Delta_t$, hence this equation is non-linear. 
It is also important to note that the pressure functional (\ref{freeenergy}) is preserved by the time evolution (\ref{eq:BdG-p}).

 In this paper we study the time evolution  (\ref{eq:BdG-p}) for initial states close to the normal state, for temperatures close to $T_c$. Closeness is measured in terms of a small parameter, which we call $h$. Let us assume that $|T-T_c| \leq h^2$, and also that the initial state $\Gamma_0$ satisfies 
 \begin{equation}\label{fh4} 
 \F(\Gamma_0) - \F(\Gamma_{\rm n}) \leq C h^4 
 \end{equation}
for some constant $C>0$ (independent of $h$)\footnote{Throughout the paper, we denote by $C$ generic constants, even if they take different values at different places.}. This is satisfied, for instance, by thermal equilibrium states in the temperature range considered, but also by states where the order parameter $\psi$ in (\ref{ap}) is modified by a factor of order one. 
In addition to the assumption $T_c>0$, we assume that $\mu>0$ and that  $\alpha_*$ does not vanish identically on the Fermi sphere, i.e.,
\begin{equation}\label{eq:ass}
 \sup_{k^2 = \mu} |\hat{\alpha}_* (\k)| > 0,
\end{equation}
 which is satisfied generically.

With $\alpha_t$ denoting the pairing density of $\Gamma_t$ we define, for every time $t$, the complex number  
\[ \psi_t = h^{-1} \langle \alpha_*| \alpha_t \rangle  . \]
For convenience, we multiply by $h^{-1}$ in order for $\psi$ to be of order one. Our main result states that, for small $h$,  $|\psi_t|$ remains approximately constant, uniformly in time.

\begin{theorem}\label{thm}
Let $\Gamma_t$ be the solution of (\ref{eq:BdG-p}) with initial state $\Gamma_0$ satisfying (\ref{fh4}), and $|T-T_c|\leq h^2$. 
Then there exists a constant $C > 0$ such that, for small $h$, 
\begin{equation}\label{claim}
 \left| |\psi_t|^2 - |\psi_0|^2 \right| \leq C h^{1/2} 
\end{equation}
for all times $t$.
\end{theorem}

We remark that the conditions of the theorem allow $|\psi_0|$ to take any value of order one. The result states that this value remains constant to leading order in $h$. This holds true even for $T> T_c$, in which case the normal state $\Gamma_{\rm n}$ minimizes the pressure functional (\ref{freeenergy}). In other words, the order parameter does not tend towards the minimum of the GL energy (\ref{eq:GL}), as would be expected on the basis of a time-dependent GL equation of the form \cite{melo,randeria}
\begin{equation}\label{eq:tGL}
i d \partial_t \psi = a \psi + b |\psi|^2 \psi 
\end{equation}
with $a\in \R$, $b>0$ and $d\in \C$ with $\Im d > 0$. In fact,  for $T>T_c$ one has $a>0$, in which case the solution to \eqref{eq:tGL} goes to zero as $t\to \infty$, in contrast to our main result \eqref{claim}. 
Moreover, at $T-T_c < 0$ (but small), one could for instance start with a state with the \lq\lq wrong\rq\rq\ $\psi$, i.e., with $\alpha(\k)$ the equilibrium pairing density multiplied by a complex number of modulus not equal to one, and our theorem states that this structure will be preserved at all times.

We emphasize that our results do not rule out the validity of the time-dependent
Ginzburg--Landau equation, in general, which has been successfully employed over
several decades. What they show, however, is that such an equation cannot be 
derived from the Bogoliubov--de-Gennes
equations, which also appear prominently in the physics  literature. From the point of view of mathematical physics, it thus remains a
challenging open problem to unveil the relevant additional physical
effects which are responsible for the possible emergence of a time-dependent GL equation.

\begin{proof}
The proof of Theorem \ref{thm} is divided into three steps.

{\it Step 1.} The first step is to show that the energy bound $\F(\Gamma) - \F(\Gamma_{\rm n}) \leq C h^4$ implies a decomposition of the form 
$$\gamma (\k) = \gamma_{\rm n} (\k) + \eta (\k) \ , \quad  \hat{\alpha} (\k) = h \psi \hat{\alpha}_* (\k) + \xi (\k) $$
where $\psi = h^{-1} \langle \alpha_* | \alpha \rangle$ and where
\begin{equation}\label{eq:apri}
 |\psi | \leq C, \quad \| \xi \|_2 \leq Ch^2 , \quad \| \eta \|_2 \leq Ch^2 
\end{equation}
for an appropriate constant $C > 0$.  
These bounds follow from the analysis of \cite{FHSS};  we shall sketch the main ideas here. From the bound  \cite[Lemma~1]{FHSS} on the relative entropy of $\Gamma$ with respect to  $\Gamma_{\rm n}$ we deduce that
  $ \F(\Gamma) - \F(\Gamma_{\rm n})$ can be bounded from below by 
 \begin{align}\label{apri}
& \langle \alpha| (K_T + V) \alpha \rangle+ \int K_T (\k) (\gamma (\k) - \gamma_n(\k))^2 d^3k \\  \nonumber & +  \frac{2T}{3} \int \! \Tr_{\C^2} \! \left[  \Gamma(\k)(1-\Gamma(\k)) - \Gamma_{\rm n}(\k)(1-\Gamma_n(\k)) \right]^2  d^3k.
\end{align}
For $T\geq T_c$, we can bound $K_T \geq K_{T_c}$ in the first term. Since  $\alpha_\ast$ is, by definition, the non-degenerate 
  zero eigenvector of $K_{T_c} + V$, with a spectral gap $\kappa>0$, we conclude that the first term in  (\ref{apri}) is  bounded from below by
$  \langle \xi | (K_{T_c} + V) \xi \rangle  \geq \kappa \|\xi \|_2^2 $ in this case. Moreover, the second term is bounded from below by $2 T \|\gamma - \gamma_{\rm n}\|^2_2 $ 
since $K_T \geq 2T$. Hence we obtain  the second and third bound in (\ref{eq:apri}). 
Moreover, with the aid of the last term in (\ref{apri}) it is not difficult to show that $|\psi| \leq C$, concluding the proof of (\ref{eq:apri}) for $T\geq T_c$. The case $T< T_c$ is very similar, using the fact that $K_T \geq K_{T_c} + 2(T- T_c) \geq K_{T_c} -2 h^2$ instead, and we refer to \cite{FHSS} for the details.

{\it Step 2.} Since the dynamics (\ref{eq:BdG-p}) is unitary, the eigenvalues of the $2 \times 2$ matrix $\Gamma_t (\k)$ are conserved, for all $\k \in \R^3$. A simple computation shows that they 
are of the form $1/2 \pm s_t(\k)$, with
\begin{equation}\label{conss}
 s_t(\k) = \sqrt{\left(\gamma_t(\k) - \tfrac 12\right)^2 + |\hat \alpha_t(\k)|^2}.
\end{equation}
That is, also $s_t(\k)$ is independent of $t$. 

{\it Step 3.}  Let $\Gamma_0$ be  an initial state satisfying  (\ref{fh4}) and let $\Gamma_t$ be the solution of (\ref{eq:BdG-p}). Since the pressure functional is conserved, we also have  $\F(\Gamma_t) - \F(\Gamma_{\rm n}) \leq C h^4$ for all $t$. Setting, as above, 
 $\psi_t = h^{-1} \langle \alpha_*| \alpha_t \rangle$, $\hat{\alpha}_t (\k) = h \psi_t \hat{\alpha}_* (\k)+ \xi_t (\k)$ and 
$\gamma_t (\k) = \gamma_{\rm n} (\k)+ \eta_t (\k)$, we find that (\ref{eq:apri}) holds for $(\psi_t,\xi_t,\eta_t)$, uniformly in $t$.

The equation $s_t(\k)^2 = s_0(\k)^2$ can be written as 
\begin{align}\nonumber
& \eta_t(\k)^2 - \eta_0(\k)^2 - \left( \eta_t(\k) - \eta_0(\k) \right)\tanh \left( \tfrac{ k^2 -\mu}{2T} \right) \\ & = |\hat\alpha_0(\k)|^2 - |\hat\alpha_t(\k)|^2  \label{10}
\end{align}
using  $1- 2  \gamma_{\rm n} (\k)  =  \tanh ((k^2 - \mu)/2T)$. We integrate this identity over a thin annulus of thickness $\delta$ around the Fermi sphere, denoted by $$\Omega_\delta = \{ \k \in \R^3 : \left| |\k| - \sqrt{\mu} \right| \leq \delta \}.$$ Since $\tanh ((k^2 -\mu)/2T) = 0$ on the Fermi sphere, 
\[ \int_{\Omega_\delta}  \tanh^2 \left( \tfrac{k^2 - \mu}{2T} \right)  d^3k \leq C \delta^3, \]
and hence (\ref{eq:apri}), together with the Cauchy-Schwarz inequality, implies that the left side of (\ref{10}) is bounded, after integration over $\Omega_\delta$, by $C(h^4 + h^2 \delta^{3/2})$. To estimate the right side, we bound
\begin{align}\nonumber
 & \left| |h \psi_t \hat{\alpha}_* (\k) + \xi_t (\k)|^2 - |h \psi_0 \hat{\alpha}_* (\k) + \xi_0 (\k)|^2 \right| \\  \nonumber & \geq  h^2 |\hat{\alpha}_\ast (\k)|^2 \left| |\psi_t|^2 - |\psi_0|^2 \right| \\ \nonumber & \quad - 2h  |\hat{\alpha}_* (\k)| \left( |\psi_t|   |\xi_t (\k)| + |\psi_0|  |\xi_0(\k)| \right) \\ & \quad - |\xi_t (\k)|^2 - |\xi_0 (\k)|^2.  \label{11}
\end{align}
The assumption (\ref{eq:ass}), together with the continuity of $\hat{\alpha}_*$ (which follows from Prop.~2 in \cite{FHSS}, since the latter implies $\alpha_*\in L^1(\R^3)$),  implies that for small $\delta$  
\[ \int_{\Omega_\delta} |\hat{\alpha}_* (\k)|^2 d^3k \geq C \delta. \]
After integration over $\Omega_\delta$, the right side of (\ref{11}) is thus  bounded from below by 
$
C h^2 \delta \left| |\psi_t|^2 - |\psi_0|^2 \right| -  C h^3 \delta^{1/2} - C h^4 
$
using again the Cauchy-Schwarz inequality and (\ref{eq:apri}). Together with the bound on the left side above, this implies that
\[ C h^2 \delta \left| |\psi_t|^2 - |\psi_0|^2 \right| \leq  C \left( h^4 + h^3 \delta^{1/2} +  h^2 \delta^{3/2}\right) .\]
The choice $\delta = h$ leads to the claim (\ref{claim}). 
\end{proof}

\section{Comparison with Linear Case}

It is interesting to observe that our main result, namely the fact that $|\psi_t|$ remains approximately constant in time, crucially depends on the nonlinear terms in the time-dependent BCS equation. Let us explain this point in more detail.  The equation for $\alpha_t$ in (\ref{eq:BdG-p}) is given by
$$
i \partial_t \hat\alpha_t(\k) = 2 (k^2-\mu) \hat\alpha_t(\k) + \Delta_t(\k) \left( 1- \gamma_t(\k) - \gamma_t(-\k) \right).
$$
Writing again $\gamma_t(\k) = \gamma_{\rm n}(\k) + \eta_t(\k)$ this can be rewritten abstractly as
\begin{equation}\label{abstra}
i \partial_t \alpha_t = L S \alpha_t  - 2(\eta_t + \eta_t^\dagger) V \alpha_t,
\end{equation}
where $S$ denotes the operator $K_T (-i\nabla) + V(\x)$, $L$ is multiplication by $2-4\gamma_{\rm n}(\k) = 2\tanh((k^2-\mu)/2T)$, $V$ is multiplication by $V(\x)$ in $\x$-space and $\eta_t$ is multiplication by $\eta_t(\k)$ in $\k$-space. 
Consider, for simplicity, the case $T > T_c$; in this case,  the operator $S$ is positive and 
the solution to (\ref{abstra}) satisfies
\begin{equation} 
\alpha_t  = S^{-1/2} U(t) S^{1/2} \alpha_0 + 2 i  \int_0^t S^{-1/2}  U(t-s) S^{1/2} (\eta_s+\eta_s^\dagger) V \alpha_s ds  \label{alpequ} 
\end{equation}
for $t>0$, where 
\begin{equation}\label{def:U}
 U(t) = e^{-i t S^{1/2} L S^{1/2}}
 \end{equation}
  is the unitary evolution generated by $S^{1/2} L S^{1/2}$. 

In the second term on the right side of (\ref{alpequ}), we can use (\ref{10}) to express $\eta_s$ in terms of $\alpha_s$; this leads to a nonlinear equation for $\alpha_t$. 
 Let us neglect for a moment this second term, and let us focus on the linear dynamics $S^{-1/2} U (t) S^{1/2} \alpha_0$. The spectrum of $S^{1/2} L S^{1/2}$ coincides with the one of $LS$. Its continuous spectrum can easily be seen to cover the halfline $[-2 \mu ; \infty)$, since $LS - 2(k^2 - \mu)$ is relatively compact with respect to $k^2$. Moreover, for $T=T_c$, $LS$ has an eigenvalue $0$ associated with the eigenvector $\alpha_*$, which is  embedded in the continuous spectrum. Perturbation theory predicts that  for $T>T_c$ the zero eigenvalue turns into a complex resonance $\lambda$, with real and imaginary parts  of the order $T-T_c$. 

It is particularly simple to find the resonance $\lambda$ of $S^{1/2} L S^{1/2}$ if the potential $V$ is rank one, i.e., of the form $V = - |\ph \rangle \langle \ph|$ for a  $\ph \in L^2 (\R^3)$ which we assume to be radial for simplicity. In this case, $T_c$ is determined by 
$ \langle \ph, K_{T_c}^{-1} \ph \rangle=1$  and  $\alpha_*$ is proportional to $K_{T_c}^{-1} \ph$. 

To compute $\lambda$, we use complex dilation. For $\theta \in \R$, we define the unitary operator $u(\theta)$ by 
\[ \left[ u (\theta) \phi \right] (\x) = e^{-3\theta/2} \phi (e^{-\theta} \x) .\] 
Alternatively, $u(\theta) = e^{i\theta A}$ with 
 $A = \x \cdot \k + \k \cdot \x$. Assuming $\ph$ to be an analytic vector for $A$, we can extend $\ph_\theta = u(\theta) \ph$ to a strip $-\beta< \im \theta \leq 0 $ below the real axis. In this way, we can also define the operators $L_\theta$ and $S_\theta$ for all $-\beta < \im \theta \leq 0$. The resonance $\lambda$ then satisfies the eigenvalue equation $L_\theta S_\theta \chi_\theta = \lambda\chi_\theta$, which is equivalent to 
$$ \chi_\theta =\frac 1{ 2 e^{-2 \theta} k^2 - 2 \mu + \lambda}  L_\theta | \phi_\theta \rangle \langle \phi_\theta  | \chi_\theta \rangle. $$
Multiplying from the left with $\langle \phi_\theta|$, we obtain 
$$1 =  \left\langle \phi_\theta \left|  \frac 1{ 2 e^{-2 \theta} k^2 - 2 \mu + \lambda}  L_\theta \right. \phi_\theta \right\rangle.$$
Note that $\lambda$ vanishes at $T= T_c$. We can use implicit differentiation with respect to $T$ to expand around $T=T_c$, letting $\theta\to 0$ afterwards. This yields
\begin{align*}\nonumber
\lambda & \simeq  \frac {T_c-T}{2T_c^2} \int |\phi(\k)|^2 \cosh^{-2}\left(\tfrac{k^2-\mu}{2T} \right) d^3k
\\ &  \quad \times 
\label{eq:lh} \left[ \text{p.v.}\! \int \frac{|\phi(\k)|^2 }{(k^2 -   \mu) K_{T_c}(\k)} d^3k - i \frac{\pi^2 \sqrt{\mu}}{T_c} |\phi(\sqrt{\mu})|^2 \right]^{-1}
\end{align*}
to leading order in $T-T_c$, where the integral in the last factor is understood in the principal value (p.v.) sense.

The fact that $\im \lambda <0 $ suggests that the corresponding state decays exponentially in time. In particular, one would expect from the linear evolution (\ref{def:U}) that the order parameter satisfies 
\begin{equation*}\label{eq:exp-dec} 
| \psi_t | \approx | \psi_0 | e^{t \im \lambda } 
\end{equation*}
to leading order in $T-T_c$, i.e., it decays on a time scale of the order $(T-T_c)^{-1}$. Such a decay was in fact predicted in \cite{Cyrot,melo,randeria}. 
 The meaning of the $\approx$ sign here can be made precise following the analysis of \cite{Hunzi}, but  the details are not relevant here. 
A comparison with the statement of Theorem \ref{thm} shows the importance of the second, nonlinear term on the right side of (\ref{alpequ}) for understanding the behavior of $\alpha_t$. Let us examine it closer. The function $\eta_t(\k)$  is determined by Eq.~(\ref{10}). Away from the Fermi sphere, the second term on the left side of (\ref{10}) dominates, and we have 
\begin{equation}\label{eq:hatVal} 
\eta_t(\k) - \eta_0(\k) \simeq \frac{ |\hat\alpha_t(\k)|^2 - |\hat\alpha_0(\k)|^2}{\tanh\left( \tfrac {k^2-\mu}{2T} \right)}
\end{equation}
to leading order, 
which leads to a cubic equation for the evolution of $\alpha_t$. In terms of $\psi_t$ this would even resemble the cubic GL term.
On the Fermi sphere the denominator on the right side  of (16) vanishes, however. As a consequence, 
$\eta_t$ is much larger, of the order  $| |\hat{\alpha}_t (\k)|^2 - |\hat{\alpha}_0 (\k)|^2|^{1/2}$ according to (\ref{10}). The latter expression equals 
$h |\hat{\alpha}_* (\k)| | |\psi_t|^2 - |\psi_0|^2|^{1/2}$ to leading order. 
Since $\hat{\alpha}_*$ does not vanish on the Fermi sphere, we conclude that $|\psi_t| \simeq |\psi_0|$; otherwise,  $\eta_t (\k)$ would be too large to satisfy (\ref{eq:apri}). This is exactly the mechanism used in the proof of Theorem \ref{thm},  explaining why the nonlinear term in (\ref{alpequ}) plays such an important role in determining the behavior of $\psi_t$.

\section*{Acknowledgments}

Financial support from the U.S. National Science Foundation through grants PHY-1347399 and DMS-1363432 (R.L.F.), SwissMAP and SNF grant Nr. 200021-153621 (B.S.), and the Austrian Science Fund (FWF) project Nr. P 27533-N27 (R.S.) is acknowledged.

\end{document}